\documentclass[journal]{IEEEtran}
\UseRawInputEncoding

\ifCLASSINFOpdf
\else
\fi
\usepackage{tabularx}
\usepackage{makecell}
\usepackage{booktabs}
\newcolumntype{P}[1]{>{\centering\arraybackslash}p{#1}}
\usepackage{dblfloatfix}
\usepackage{amssymb}
\usepackage{amsmath}
\usepackage{graphicx}
\usepackage{bm}
\usepackage[noabbrev]{cleveref}
\usepackage{todonotes}
\usepackage{tabularx}

\usepackage{sfmath}
\usepackage{soul}

\renewcommand{\arraystretch}{2}
\DeclareMathOperator*{\argmin}{\arg\!\min}
 
\let\oldhat\hat
\let\lhat\hat
\renewcommand{\hat}[1]{\skew{7}\oldhat{#1}}
\renewcommand{\lhat}[1]{\skew{-4}\oldhat{#1}}

\def\hs{\hspace{\fontdimen2\font}}
	
\begin{document}

\title{Spatial response identification enables robust experimental ultrasound computed tomography}

\author{Carlos Cueto,
        Lluis Guasch,
        Javier Cudeiro,
        \`{O}scar Calder\'{o}n Agudo,
        Oscar Bates,
        George Strong,
        and Meng-Xing Tang
\thanks{This work was supported by the Wellcome Trust 219624/Z/19/Z. The work of Carlos Cueto was supported by the EPSRC Centre for Doctoral Training in Medical Imaging under Grant EP/L015226/1. The work of Oscar Bates was supported by the EPSRC Centre for Doctoral Training in Neurotechnology under Grant EP/L016737/1.}
\thanks{Carlos Cueto, Oscar Bates and Meng-Xing Tang are with the Ultrasound
Laboratory for Imaging and Sensing Group, Department of Bioengineering,
Imperial College London, London SW7 2AZ, U.K. (e-mail: c.cueto@imperial.ac.uk, o.bates18@imperial.ac.uk, mengxing.tang@imperial.ac.uk).}
\thanks{Lluis Guasch, Javier Cudeiro, \`{O}scar Calder\'{o}n Agudo, and George Strong are with the Department of Earth Science and Engineering, Imperial College London, London SW7 2AZ, U.K. (e-mail: l.guasch08@imperial.ac.uk, j.cudeiro15@imperial.ac.uk, o.calderon-agudo14@imperial.ac.uk, george.stronge14@imperial.ac.uk).}
}

%



\maketitle

\begin{abstract}

Ultrasound computed tomography techniques have the potential to provide clinicians with 3D, quantitative and high-resolution information of both soft and hard tissues such as the breast or the adult human brain. Their practical application requires accurate modelling of the acquisition setup: the spatial location, orientation, and impulse response of each ultrasound transducer. However, existing calibration methods fail to accurately characterise these transducers unless their size can be considered negligible when compared to the dominant wavelength, which reduces signal-to-noise ratios below usable levels in the presence of high-contrast tissues such as the skull. In this paper, we introduce a methodology that can simultaneously estimate the location, orientation, and impulse response of the ultrasound transducers in a single calibration. We do this by extending spatial response identification, an algorithm that we have recently proposed to estimate transducer impulse responses. Our proposed methodology replaces the transducers in the acquisition device with a surrogate model whose effective response matches the experimental data by fitting a numerical model of wave propagation. This results in a flexible and robust calibration procedure that can accurately predict the behaviour of the ultrasound acquisition device without ever having to know where the real transducers are or their individual impulse response. Experimental results using a ring acquisition system show that spatial response identification produces calibrations of significantly higher quality than standard methodologies across all transducers, both in transmission and in reception. Experimental full-waveform inversion reconstructions of a tissue-mimicking phantom demonstrate that spatial response identification generates more accurate reconstructions than those produced with standard calibration techniques.

\end{abstract}


%
\IEEEpeerreviewmaketitle

\section{Introduction}

Ultrasound computed tomography techniques such as full-waveform inversion (FWI) are emerging as an effective tool to quantitatively image human tissue in 3D, with applications in breast \cite{Wiskin20173-DResults,Sandhu2015FrequencyTransducer}, limb \cite{Wiskin2020FullContrast} and brain \cite{Guasch2020Full-waveformBrain} imaging. The successful application of these techniques revolves around reliable wave-propagation modelling, which requires that the spatial location and orientation of sources and receivers, as well as their impulse response (IR), are known with high accuracy. However, at present these can only be estimated accurately in a limited number of cases, hindering the application of these tomographic techniques under experimental conditions.

The spatial location of sources and receivers could be determined from the nominal geometry of the acquisition device, but this generally lacks sufficient accuracy and locations are usually estimated using time-of-flight (TOF) based calibrations \cite{Filipik2008ModifiedTomography,Roy2011RobustTomography}. These methods use the time taken by ultrasound signals to travel between pairs of sources and receivers in the device to triangulate their location. Despite their widespread use in ultrasound computed tomography, TOF-based methods fail to recover accurate locations when the surface of transducers cannot be considered negligibly small with respect to the dominant wavelength, and they are unable to account for transducer orientation. Although transducers with small surfaces have been used to work around these issues in prior studies \cite{Wiskin20173-DResults,Sandhu2015FrequencyTransducer,CalderonAgudo20183DInversion}, their smaller active surface results in lower signal-to-noise ratios that are especially relevant when imaging high-contrast targets such as the head \cite{Guasch2020Full-waveformBrain}.

On the other hand, a number of methods exist to estimate transducer IRs in transmission and reception, with deconvolution-based methods \cite{Jensen2016SafetySimulations,Warner2013AnisotropicInversion} and backpropagation \cite{Sapozhnikov2006TransientTransducer,Sapozhnikov2015AcousticFields,Alles2011IterativeCharacterization,Treeby2018Equivalent-SourceMedia} among the most used ones. Deconvolution-based methods assume that the transducer behaves like a baffled piston and calculate an IR by deconvolving the expected infinite-bandwidth response of the transducer and its experimentally measured response. In backpropagation, which is only applicable in transmission, the response of the transducer is reconstructed by running a model of wave propagation backwards in time until signals recorded experimentally with a hydrophone are projected onto the surface of the transducer. As we showed in previous work \cite{Cueto2021SpatialImaging}, both of these methods fail to achieve the accuracy that is required by ultrasound tomography techniques such as FWI. This led us to propose spatial response identification (SRI), an impulse response estimation algorithm that achieves higher experimental calibration performance than existing techniques.

In this work, we extend our formulation of SRI to account not only for the IR of the transducers, but also for the location and orientation of sources and receivers within the acquisition device. Our proposed approach offers a flexible and accurate calibration methodology that absorbs these causes of experimental uncertainty while simultaneously honouring the full physics of the wave equation. Based on this extended formulation, we show that the application of SRI produces improved experimental FWI reconstructions of a tissue-mimicking phantom when compared to standard calibration techniques.

The Methods section introduces the formulation of our proposed SRI, followed by a description of the ultrasound acquisition system and the calibration setup. Lastly, it introduces the tissue-mimicking phantom and the tomographic setup used to image it. In the Results section, we compare the performance of SRI and the above-mentioned standard techniques in terms of their ability to accurately describe the acquisition system. This is followed by results from FWI reconstructions of the tissue-mimicking phantom. Finally, we proceed to our Discussion and present our Conclusions.

\section{Methods}

\subsection{Spatial response identification}

\begin{figure*}[!t]
    \centering
    \includegraphics{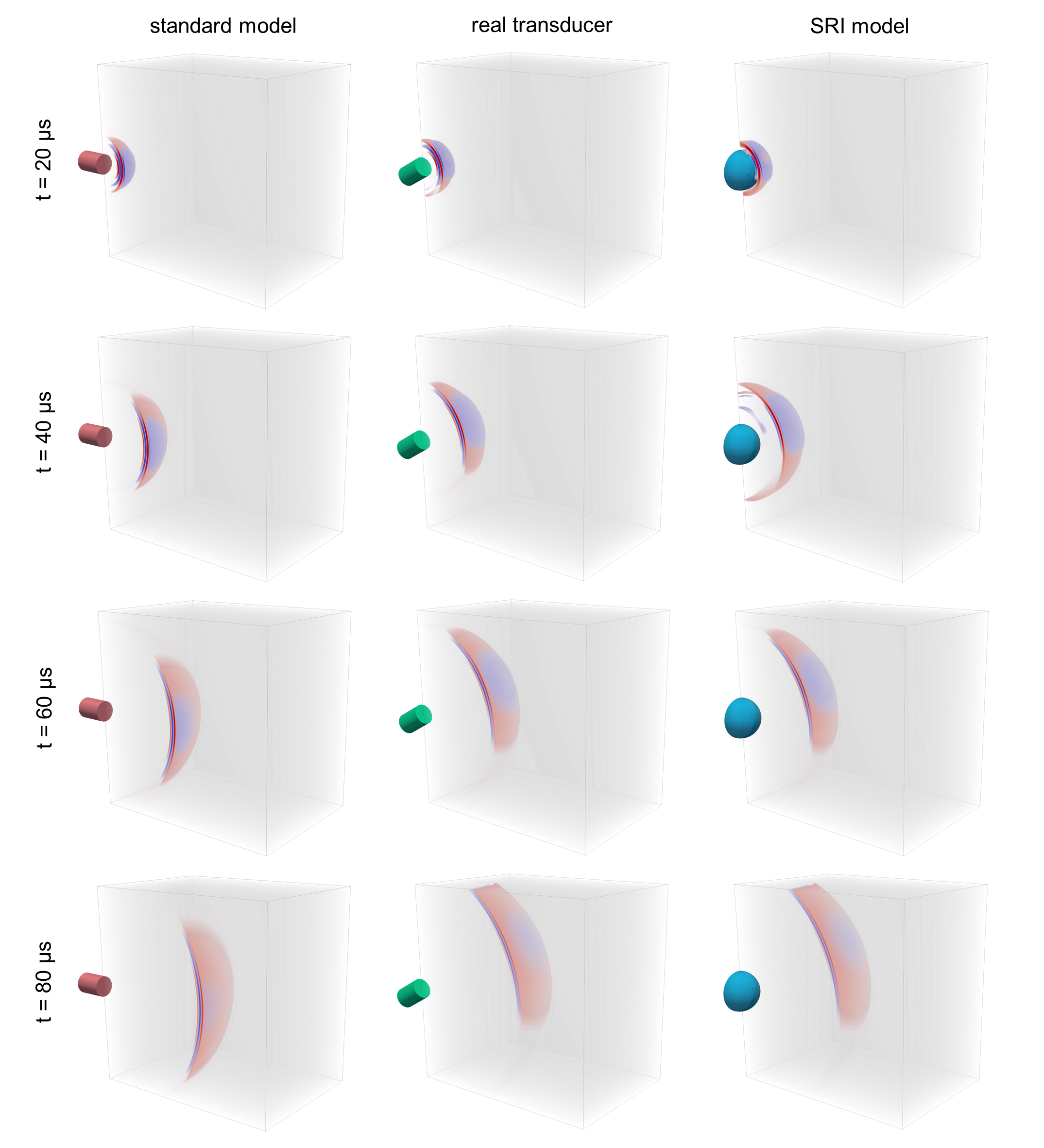}
    \caption{Illustration of the proposed SRI methodology. The goal of a calibration is to emulate the experimental behaviour of real ultrasound transducers with an unknown location, orientation and impulse response (middle column). With standard calibration techniques, we can approximate the behaviour of the real transducer, but this usually results in inaccuracies due to the finite-sized nature of the device (left column). Alternatively, we introduce a surrogate model, a hemisphere surrounding the nominal location of the transducer, and optimise the phase and amplitude emitted by each point on that surface to match the behaviour of the real transducer (right column).}
    \label{fig:wavefields}
\end{figure*}

\begin{figure*}[!b]
    \centering
    \includegraphics{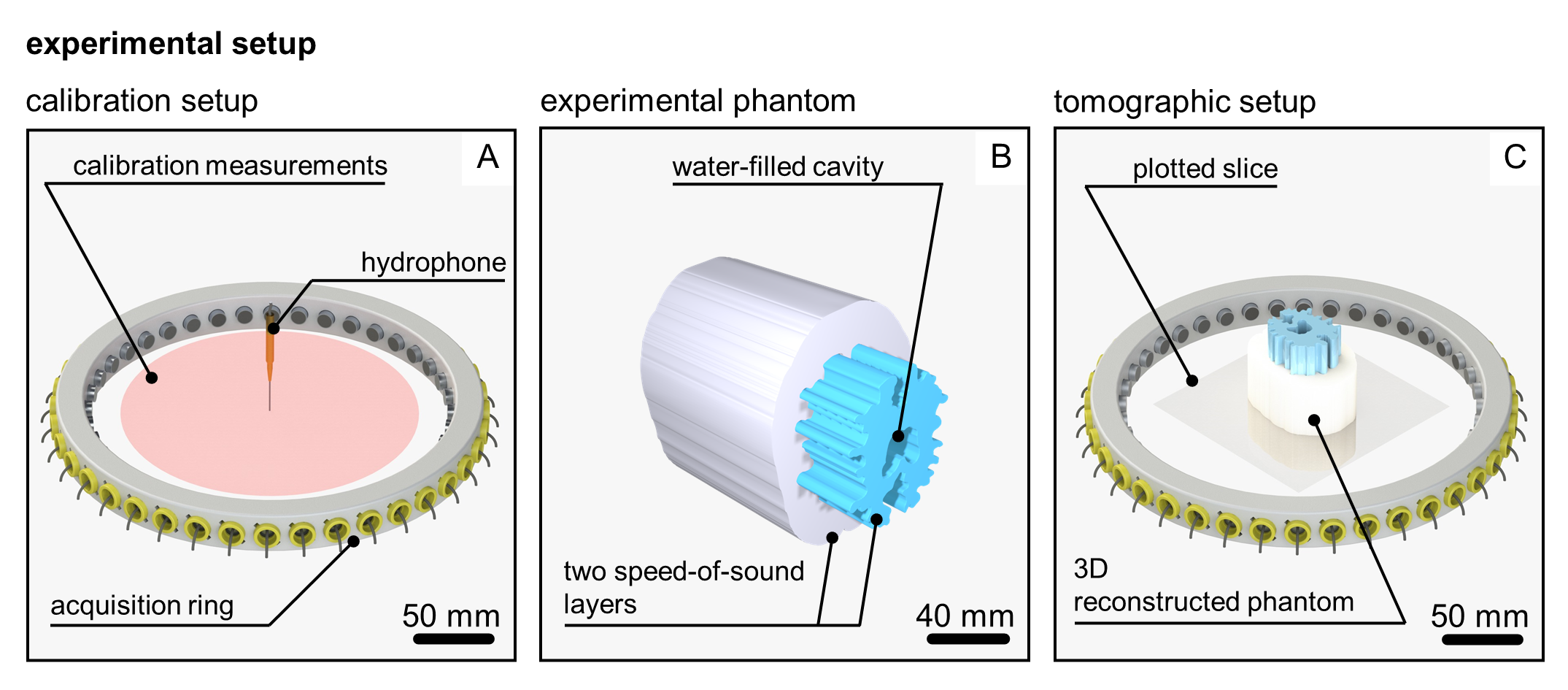}
    \caption{Experimental setup and imaging phantom. A ring acquisition system with 40 disk transducers was calibrated by introducing a needle hydrophone vertically intro the system and automatically scanning the area represented as a red disk (A). A tissue-mimicking phantom with two layers of different speed of sound values and an inner cavity filled with water was used as an imaging target after the calibration (B). The phantom was introduced vertically into the centre of the ring and scanned as the ring rotated ten times in steps of 1 degree; the reconstruction was carried out in 3D and the central slice was selected for plotting purposes (C).}
    \label{fig:experimental_setup}
\end{figure*}

Robust ultrasound computed tomography requires accurate modelling of the experimental acquisition device, both in transmission and in reception. This requirement stems from the fact that ultrasound tomography algorithms rely on high-fidelity models of wave propagation. An accurate model of the acquisition device needs to account for both the spatial location and orientation of the transducers in it and for their IRs. In practice, that entails being able to accurately reproduce the acoustic wavefield generated by every source in the device when excited by an electric signal, as well as being able to estimate the output signal generated by every receiver when exposed to an impinging acoustic wave.

We can take the middle column in Fig. \ref{fig:wavefields} as an example of a device whose response we want to calibrate in the transmission regime. A source with a certain location, orientation and IR is emitting a pressure field that we would like to model accurately. Through standard calibration methods, we could naively estimate the location and IR of the transducer, but these calibrations will suffer from inaccuracies due to the finite size of the transducer, as can be seen in the left column of Fig. \ref{fig:wavefields}.

Instead, we surround the area where the real transducer is nominally located with a hemispherical surface, and we manipulate the amplitude and phase of the wave emitted from each point on this surface such that the generated pressure field approximates that of the real transducer without ever having to know its actual location, orientation or IR, as seen in the right column of Fig. \ref{fig:wavefields}. The calibration problem then becomes that of finding the optimal amplitude and phase manipulations required to reproduce the behaviour seen experimentally for the real transducer.

As illustrated by the example presented above, our proposed calibration methodology starts by surrounding the nominal location of each transducer on the acquisition device with a hemispherical surface acting as a surrogate model. This surface is assigned a spatially varying impulse response that is different in transmission and in reception. Then, we use SRI, an impulse response estimation algorithm, to calibrate the surrogate model so that it matches the experimental data while honouring the physics of wave propagation.

By recasting the problem in this way, the hemispherical surface becomes a black box that can simultaneously absorb the location, orientation and physical impulse response of the transducer by optimising the spatially varying behaviour of the surrogate model. This reformulation of the problem is possible because we are not really interested in knowing the characteristics of the real transducer, but only its effective response.

In the following paragraphs, we review the formulation of SRI as presented in \cite{Cueto2021SpatialImaging} for completeness. SRI assumes that the arbitrary transducer surface has a spatially varying response that can be parametrised as an IR and its effect is that of a convolution operation. In transmission, the IR $\mathbf{h}_T(t, \mathbf{x})$ establishes a map from the electric voltage $\mathbf{v}_T(t)$ used to excite the transducer, to the mechanical perturbation $\mathbf{q}(t, \mathbf{x})$ generated at the surface of the transducer, which could both represent the acceleration of the medium surrounding the transducer with respect to its equilibrium state, or the time rate at which mass is injected into that medium. In reception, the IR $\mathbf{h}_R(t, \mathbf{x})$ represents the transformation of a mechanical wave $\mathbf{u}(t, \mathbf{x})$ travelling through the medium into an outgoing electric voltage $\mathbf{v}_R(t)$. Both impulse responses vary over the surface of the surrogate model as described by the spatial vector $\mathbf{x}$, and these transductions can be expressed as,

\begin{equation}
    \mathbf{q}(t, \mathbf{x}) = \mathbf{h}_{T}(t, \mathbf{x}) * \mathbf{v}_T(t) = \int_{t_0}^{t_1} \mathbf{h}_{T}(t-\tau, \mathbf{x}) \circ \mathbf{v}_T(\tau) d\tau
\label{eq:param_transmission}
\end{equation}

\begin{equation}
\begin{split}
    \mathbf{v}_R(t) &= \int_\Omega \mathbf{h}_{R}(t, \mathbf{x}) * \mathbf{u}(t, \mathbf{x}) d\mathbf{x}^3 \\
    &= \int_{t_0}^{t_1} \int_\Omega \mathbf{h}_{R}(t-\tau, \mathbf{x}) \circ \mathbf{u}(\tau, \mathbf{x}) d\mathbf{x}^3 d\tau
\end{split}
\label{eq:param_reception}
\end{equation}

\noindent
where $*$ represents the temporal convolution operator limited to the time interval $t \in [t_0, t_1]$ during which the state of the system is different from equilibrium, $\Omega \in \mathbb{R}^{3}$ represents the bounded spatial domain where the transducer is defined, and $\circ$ represents the Hadamard or element-wise vector product.

This transduction model is completed with a partial differential equation (PDE) that describes how the mechanical perturbations $\mathbf{q}(t, \mathbf{x})$ excited in transmission generate mechanical waves $\mathbf{u}(t, \mathbf{x})$ that travel through the medium until they reach the receiving transducer,

\begin{equation}
    \mathbf{L}(\mathbf{u};\mathbf{m},\mathbf{s}, t,\mathbf{x}) =  \lhat{\mathbf{L}}(\mathbf{u};\mathbf{m},\mathbf{s} , t, \mathbf{x}) - \mathbf{q}(t, \mathbf{x}) = \mathbf{0}
\label{eq:param_pde}
\end{equation}

\noindent
where $\lhat{\mathbf{L}}(\mathbf{u};\mathbf{m},\mathbf{s} , t, \mathbf{x})$ represents some linear or non-linear differential operator on the wavefield $\mathbf{u}$, which depends on the set of transducer parameters $\mathbf{m} = (\mathbf{h}_T, \mathbf{h}_R)$ and the set of medium parameters $\mathbf{s}$. Here, the dependency of the fields on the parameters and the dependency of the fields and parameters on space and time are assumed implicit.

Based on this parametrisation, finding the specific set of parameters $\mathbf{m} = (\mathbf{h}_T, \mathbf{h}_R)$ can be expressed as an optimisation problem, in which the optimal impulse response is that, which minimises a function measuring the misfit between data that has been measured experimentally (observed data) and data that is predicted by our transducer model (predicted data),  

\begin{equation}
\begin{split}
    \mathbf{m}^* = \argmin_{\mathbf{m}} J(\mathbf{u}, \mathbf{m}) \\
    s.t.\; \mathbf{L}(\mathbf{u};\mathbf{m}, t, \mathbf{x}) = \mathbf{0}
\end{split}
\label{eq:opt_1}
\end{equation}

\noindent
where $J(\mathbf{u}, \mathbf{m})$ is a scalar cost function or functional, and the problem is subject to the PDE. This optimisation problem can be solved efficiently by using an adjoint gradient formulation for any cost function, any relevant PDE, and any arbitrary propagation medium. For a derivation of an adjoint solution for the L2-norm of the misfit between predicted and observed data and the second-order acoustic wave equation see \cite{Cueto2021SpatialImaging}.

Therefore, our proposed method is based on, firstly, substituting the acquisition transducers with a series of surrogate models, composed of spatially varying convolutional kernels. Then, SRI is used to estimate the optimal set of kernels that can explain the experimental data in transmission and in reception. These spatially varying IRs act as a proxy that establishes a map between transducer input and experimentally measured output. The physical details of spatial location and orientation, as well as impulse response, are thus implicitly estimated, but never explicitly modelled.

\subsection{Calibration setup}

We validated the proposed method by experimentally calibrating a tomographic ring acquisition system (Fig. \ref{fig:experimental_setup}-A). The ring is 250 mm in diameter and contains 40 uniformly distributed disk transducers of 10 mm in diameter (Blatek Industries). The ring can rotate around its geometrical centre to acquire complementary views of a scanned object. The transducers were excited in all experiments with a one-cycle tone burst centred at 450 kHz.

Two calibrations are needed to fully model the ring behaviour, one in transmission and one in reception. For the transmission calibration, a needle hydrophone of 0.5 mm diameter (Precision Acoustics) was introduced vertically into the ring as seen in Fig. \ref{fig:experimental_setup}-A and an automatic system was used to scan the wavefield generated by each of the transducers as they acted as sources. A total of 697 measurements were performed for each of the 40 sources. They were acquired over a disk (186 mm diameter) at the geometric centre of the ring, represented as a red surface on Fig. \ref{fig:experimental_setup}-A, which was uniformly sampled over a Cartesian grid with a spacing of 6 mm.

For the reception calibration, the needle hydrophone was removed and data were recorded as one transducer acted as a source while the 35 transducers located in front of it acted as receivers. The next transducer in the ring was then selected as a source, with the 35 in front of it acting as receivers, and the process was repeated until all 40 transducers had been used. Five transducers were excluded from each of these acquisitions (the emitting one plus two on either side of it) because they did not contain usable signals for the calibration.

Acquired data were averaged over 50 repeats, processed to remove electronic noise before the arrival of the transmitted wave, and band-pass filtered in the frequency range 100-900 kHz. All experiments were carried out in a water bath, for which the speed of sound, considered homogeneous, was estimated through independent time-of-flight ultrasound measurements. These measurements were acquired with the 0.5 mm needle hydrophone as it was displaced on a known spatial grid (50 mm x 10 mm in size with a spacing of 10 mm).

For the SRI calibration, we considered each transducer within the acquisition device to have its own pair of transmission and reception IRs. The hemispheres were chosen to be 20 mm in diameter and were centred around the nominal location of the transducers. The hemispherical surfaces were discretised using 200 spatially distributed points, with each spatial point representing an IR filter of 3500 time steps. The number of points was chosen to ensure a density of at least 4 points per squared wavelength at the centre frequency of the transducers, 450 kHz \cite{Cueto2021SpatialImaging}. The points were distributed over the surface of the hemisphere using Fibonacci spirals in order to generate an even distribution \cite{Vogel1979AHead}. 

The resulting SRI calibration was compared to both the baseline uncalibrated case, and to a standard calibration using TOF to estimate the transducer locations and using deconvolution to obtain their IRs. The hydrophone data grid acquired for the transmission calibration was processed to automatically extract TOFs using an algorithm based on the Akaike information criterion \cite{Li2009AnTomography}. These were then fed into an optimisation problem to estimate the spatial location of sources and receivers, which was solved using a trust-region reflective algorithm \cite{Branch1999SubspaceProblems}. Then, the IRs were estimated, in transmission and reception, from the same datasets used in the SRI case. This was performed by assuming that the transducers behave as baffled pistons and deconvolving the expected infinite-bandwidth response of the transducer and its experimentally measured response. Deconvolution was chosen for the comparison because, on average, it performs better than other standard calibration techniques \cite{Cueto2021SpatialImaging}.

In order to measure their performance, the uncalibrated baseline, as well as the standard and SRI calibrations were introduced into an analytic solution of the acoustic wave equation for a homogeneous propagation medium. This model was used to predict the expected response of the acquisition system, which was compared to the data measured experimentally in transmission and reception. The performance of the calibrations was measured in terms of error achieved in the prediction, which was separated into magnitude error and phase error. The magnitude error was calculated as the relative error of the magnitude in the frequency domain in the range 100 kHz to 900 kHz. The phase error was calculated as the mean of the absolute value of the cross-correlation lag with respect to the duration of one cycle at the centre frequency of 450 kHz—to prevent phase-wrapping errors.

\begin{figure*}[!t]
    \centering
    \includegraphics{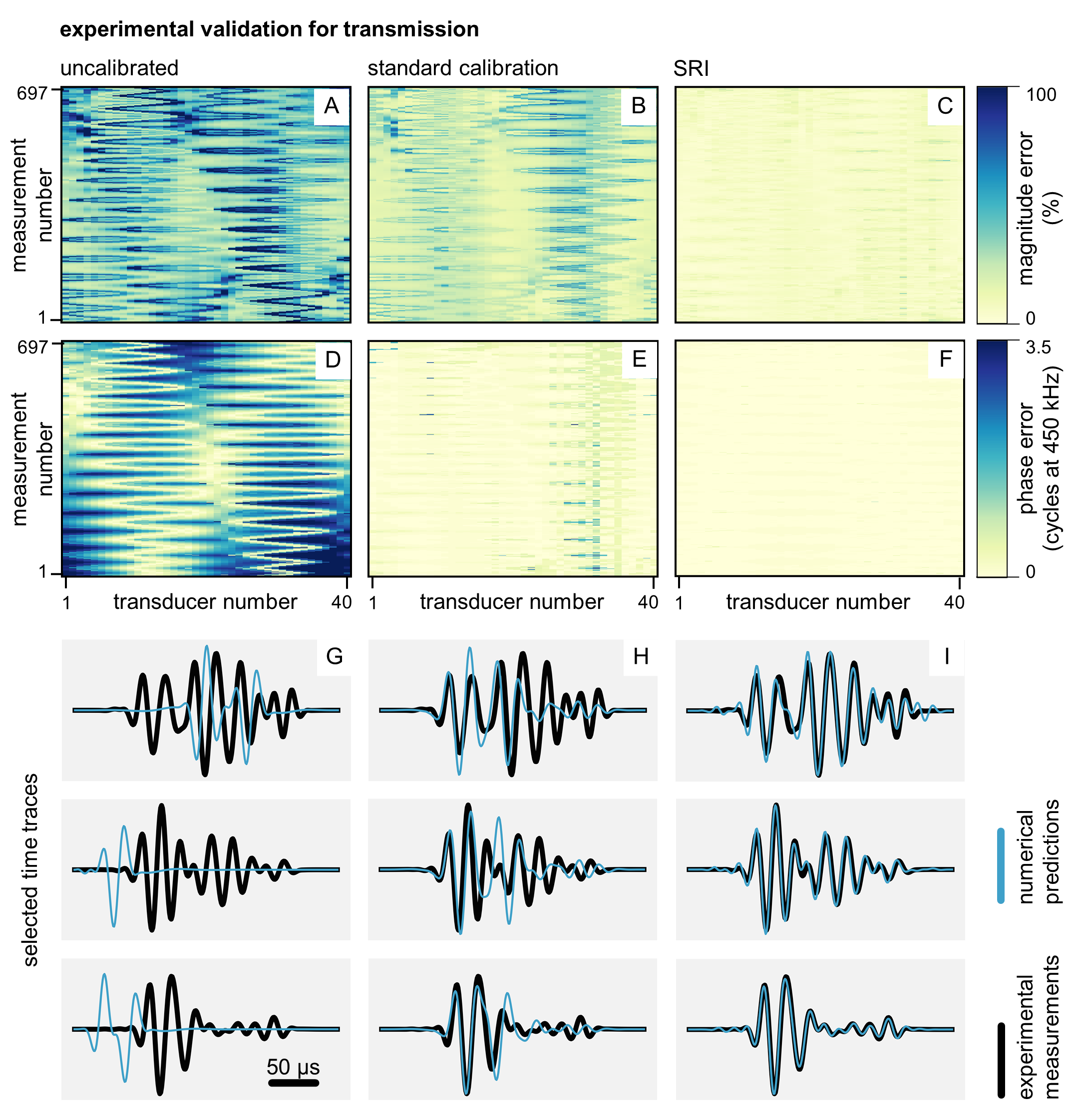}
    \caption{Performance comparison for the calibration of the ring acquisition system in transmission. The uncalibrated baseline (left column), is compared to the standard procedure combining TOF- and deconvolution-based methods (middle column), and to our proposed SRI method (right column). The performance of the calibration is evaluated in terms of magnitude error calculated in the frequency range 100 kHz to 900 kHz (A-C), and in terms of phase error expressed in number of cycles at the centre frequency of the transducers 450 kHz (D-F). Each method's predictive performance is shown for each of the total 697 experimental hydrophone measurements and for each of the 40 transducers in the ring acquisition system. Some selected time traces are also shown for the three compared cases, showcasing the match between numerical predictions and experimental measurements (G-I).}
    \label{fig:transmission_calibration}
\end{figure*}

\begin{figure*}[!t]
    \centering
    \includegraphics{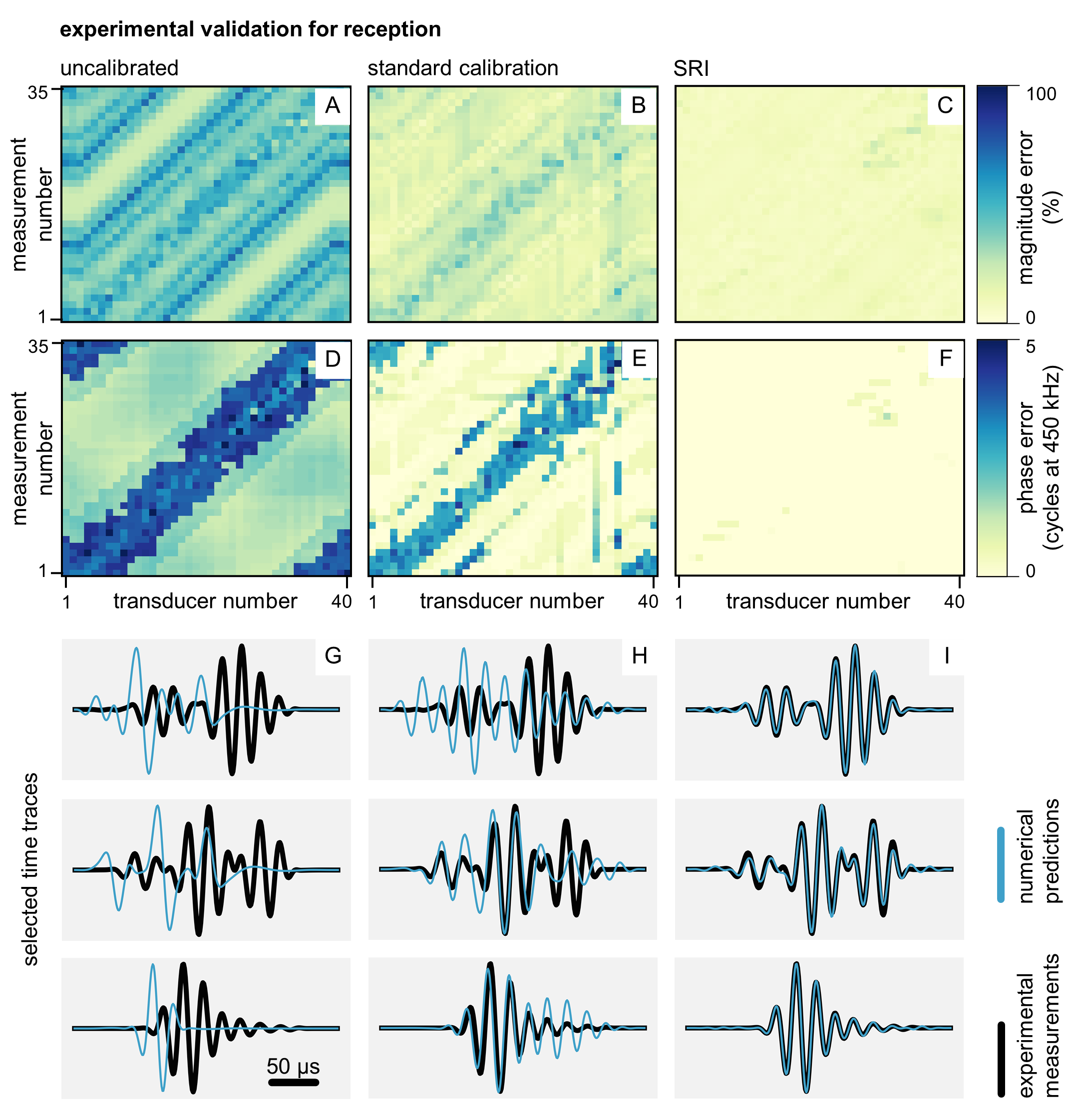}
    \caption{Performance comparison for the calibration of the ring acquisition system in reception. The uncalibrated baseline (left column), is compared to the standard procedure combining TOF- and deconvolution-based methods (middle column), and to our proposed SRI method (right column). The performance of the calibration is evaluated in terms of magnitude error calculated in the frequency range 100 kHz to 900 kHz (A-C), and in terms of phase error expressed in number of cycles at the centre frequency of the transducers 450 kHz (D-F). Each method's predictive performance is shown for each of the 35 experimental reception measurements and for each of the 40 transducers in the ring acquisition system. Some selected time traces are also shown for the three compared cases, showcasing the match between numerical predictions and experimental measurements (G-I).}
    \label{fig:reception_calibration}
\end{figure*}

\begin{figure*}[!t]
    \centering
    \includegraphics{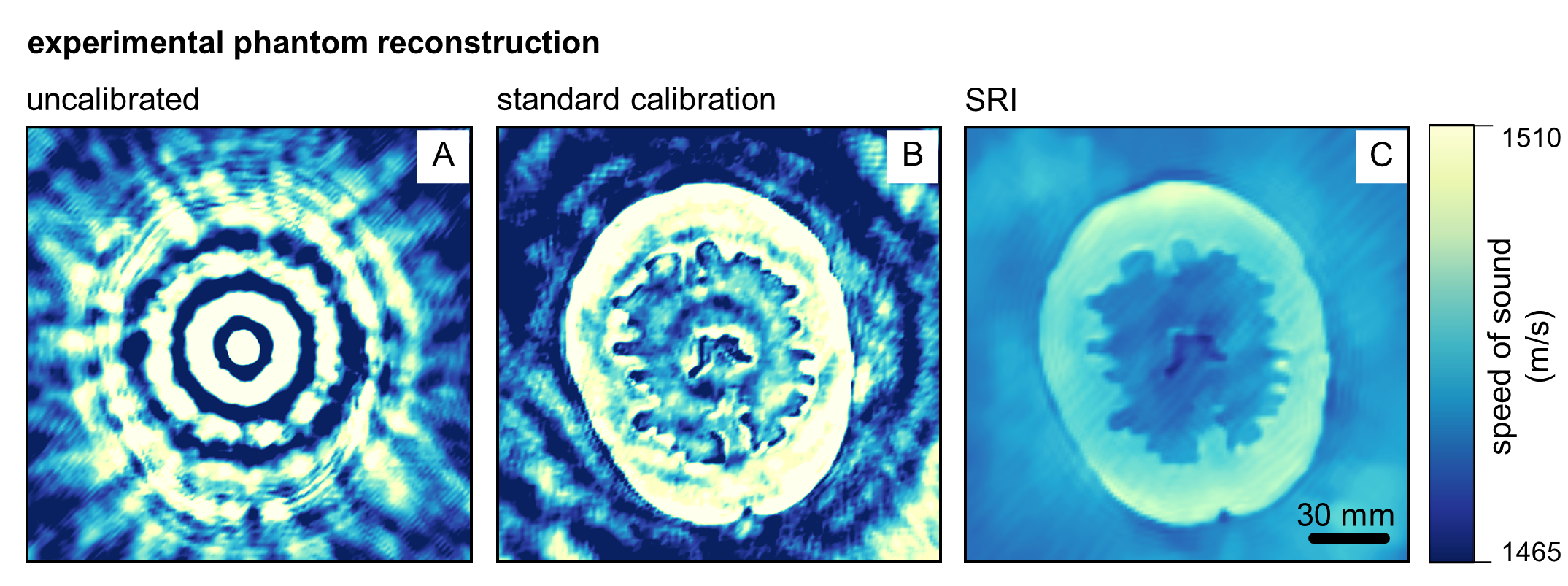}
    \caption{Impact of different calibration approaches on experimental FWI reconstruction of a tissue-mimicking phantom. When no calibration is performed, FWI fails to produce a meaningful reconstruction (A); the application of a standard calibration procedure manages to make the reconstruction feasible, but severe artefacts remain (B); the proposed SRI method results in reduced artefacts and improved reconstruction accuracy. Images show a horizontal slice through the 3D reconstruction of the phantom.}
    \label{fig:phantom_reconstruction}
\end{figure*}

\subsection{Tomographic setup}

The impact of the different calibration strategies was assessed experimentally by imaging a tissue-mimicking phantom using FWI in combination with the uncalibrated baseline model, the standard calibration and our proposed SRI model. FWI is a technique, originally developed in geophysics, that tries to recover the physical properties of the medium (usually speed of sound) by minimising the misfit between a set of experimentally measured and modelled data \cite{Virieux2009AnGeophysics,Pratt2007Sound-speedData,Guasch2020Full-waveformBrain}.

A 2.5D polyvinyl alcohol (PVA) cryogel phantom was constructed with two layers of different speed of sound values and an inner cavity filled with water as can be seen in Fig. \ref{fig:experimental_setup}-B \cite{Chee2016WalledDynamics}. A 2.5D phantom was used to prevent off-plane effects that cannot be correctly captured by a ring acquisition system. The phantom was scanned using the acquisition ring (Fig. \ref{fig:experimental_setup}-C) by emitting with each of the 40 transducers in turn while all others acted as receivers. A total of 10 rotations of one degree were performed around the phantom, resulting in 400 individual source locations. The excitation signal used was a one-cycle tone burst centred at 450 kHz. Data were processed similarly to the calibration data, by averaging over 50 repeats, removing electronic noise and band-pass filtering in the range 100-900 kHz.

The 3D inversion was carried out through 140 iterations starting at 180 kHz and progressively introducing higher frequencies up to 600 kHz. The inversion was performed in 3D because we have to account for the effect of the disk transducers in all dimensions in order to fully explain their experimental response.

In order to assess the quantitative accuracy of the reconstructions, we calculated two different metrics. First, we measured the reconstruction accuracy within the phantom by calculating the structural similarity index measure (SSIM) with respect to the computer design that we used to 3D print the phantom moulds. Second, we quantified the reconstruction precision by calculating the standard deviation (SD) of the speed of sound in each of the sub-regions of the image. Because the phantom was manufactured to have homogeneous speed of sound within each of its layers and the water regions have a homogeneous sound speed, measuring uniformity showcases the precision achieved by each reconstruction approach. To separate the different sub-regions, the computer design of the phantom was overlaid on the images and four different regions were segmented: the water bath, the inner and outer layers of the phantom, and the water-filled cavity.

\section{Results}

\subsection{Acquisition system calibration}

\begin{table*}[!b]
    \renewcommand{\arraystretch}{1.3}
    \caption{Summary performance for the ring acquisition system calibration, shown as the mean magnitude and phase errors plus one standard deviation.}
    \label{tab:calibration}
    \centering
    
    \begin{tabularx}{\textwidth}{c *{4}{P{\textwidth/5 - \tabcolsep - 0.3 mm}}}
        \hline
        
        \multicolumn{1}{c}{} & 
        \multicolumn{2}{P{\textwidth*2/5}}{{\bfseries transmission calibration}} & 
        \multicolumn{2}{P{\textwidth*2/5}}{{\bfseries reception calibration}} \\
        
        \cmidrule(lr){2-3}
        \cmidrule(lr){4-5}
        
        \multicolumn{1}{c}{} & 
        \thead{mean magnitude error \\ (\%)} & 
        \thead{mean phase error \\ (cycles at 450 kHz)} &
        \thead{mean magnitude error \\ (\%)} & 
        \thead{mean phase error \\ (cycles at 450 kHz)} \\
        
        \hline\hline
        uncalibrated         & 47.17 $\pm$\hs 33.58  & 1.48 $\pm$\hs 1.05 & 
        40.96 $\pm$\hs 11.37  & 2.30 $\pm$\hs 1.16 \\
        standard calibration & 17.28 $\pm$\hs 7.55   & 0.13 $\pm$\hs 0.29 &
        19.88 $\pm$\hs 6.71   & 0.89 $\pm$\hs 1.16 \\
        SRI                  & 5.94 $\pm$\hs 3.07    & 0.02 $\pm$\hs 0.04 & 
        8.24 $\pm$\hs 2.08    & 0.01 $\pm$\hs 0.08 \\
        \hline
    \end{tabularx}
\end{table*}

The uncalibrated baseline is compared to two different calibration approaches: (i) a standard calibration using TOF- and deconvolution-based methods, and (ii) our proposed SRI methodology. For the uncalibrated baseline, transducers were assumed to be located at nominal spatial positions, and sources and receivers were modelled as baffled pistons with an infinite bandwidth. Fig. \ref{fig:transmission_calibration} shows the results for the transmission comparison, in terms of magnitude (Fig. \ref{fig:transmission_calibration}-A to C) and phase (Fig. \ref{fig:transmission_calibration}-D to F) errors. The quality of the match for some selected time traces is also shown in Fig. \ref{fig:transmission_calibration}-G to I.

Certain spatially distributed patterns can be observed in Fig. \ref{fig:transmission_calibration}-A to F, although they are most apparent for the uncalibrated and standard cases. These patterns represent the increasing and decreasing mismatch between experimental measurements and numerical predictions as a function of the location of the hydrophone with respect to each transducer in the ring. These spatial variations showcase the fact that the emitted wavefield behaves more ideally when the hydrophone is right in front of the emitting source and moves away from the expected behaviour with increasing angle between source and hydrophone.

Similarly, Fig. \ref{fig:reception_calibration} shows the results for the reception calibration. Here, the performance is evaluated also as a function of magnitude (Fig. \ref{fig:reception_calibration}-A to C) and phase (Fig. \ref{fig:reception_calibration}-D to E) errors. Some example time traces show the quality of the calibrations with respect to experimental measurements (Fig. \ref{fig:reception_calibration}-G to I).

Also in reception, we can observe the presence of certain spatial patterns in Fig. \ref{fig:reception_calibration}-A to F, although these are also more evident for the uncalibrated and standard approaches. Once more, these show how the behaviour of a receiver being calibrated is more ideal when the source is located right in front of it and becomes more difficult to model as the angle between source and receiver increases within the ring.

Table \ref{tab:calibration} offers a summary of the calibration performance achieved by each approach. Mean and standard deviation values for the magnitude and phase errors are shown here for the uncalibrated model, the standard approach and SRI. We can see a reduction in both types of error for the SRI calibration, as well as a reduction in their spread, as shown by the decrease in the standard deviation.

Results in Fig. \ref{fig:transmission_calibration}, Fig. \ref{fig:reception_calibration} and Table \ref{tab:calibration} show that SRI outperforms standard approaches for acquisition device calibration when transducers with large surfaces are used by absorbing location, orientation and IR uncertainties into a surrogate model. We can also see how SRI has the capacity of reducing errors at all angles regardless of the calibrated transducer.

\subsection{Experimental phantom reconstruction}

\begin{table*}[!t]
    \renewcommand{\arraystretch}{1.3}
    \caption{Quantitative measures of reconstruction accuracy for the tissue-mimicking phantom. SSIM values closer to one imply higher structural accuracy, and lower SD values represent higher reconstruction precision as a measure of speed-of-sound uniformity.}
    \label{tab:phantom_reconstruction}
    \centering
    
    \begin{tabularx}{\textwidth}{c *{6}{P{\textwidth/7 - 2\tabcolsep}}}
        \hline
        &
        \thead{SSIM} & 
        \thead{mean SD \\ (m/s)} &
        \thead{water SD \\ (m/s)} &
        \thead{outer layer SD \\ (m/s)} &
        \thead{inner layer SD \\ (m/s)} &
        \thead{cavity SD \\ (m/s)} \\
        
        \hline\hline
        uncalibrated         & 0.24 & 47.60 & 15.18 & 22.91 & 48.82 & 103.51 \\
        standard calibration & 0.71 & 36.83 & 15.54 & 16.09 & 12.16 & 22.60  \\
        SRI                  & 0.85 & 3.49  & 2.36  & 4.26  & 3.40  & 3.93   \\
        \hline
    \end{tabularx}
\end{table*}

Results for the phantom reconstructions are shown in Fig. \ref{fig:phantom_reconstruction} for each of the three cases considered in the calibration validation: (i) the uncalibrated baseline (Fig. \ref{fig:phantom_reconstruction}-A), (ii) the standard approach using TOF- and deconvolution-based methods (Fig. \ref{fig:phantom_reconstruction}-B), and (iii) our proposed SRI approach (Fig. \ref{fig:phantom_reconstruction}-C). The FWI reconstruction of the phantom was performed in 3D, from which the central slice was selected for plotting purposes (Fig. \ref{fig:experimental_setup}-C).

As expected, the uncalibrated baseline (Fig. \ref{fig:phantom_reconstruction}-A) fails to produce an adequate reconstruction of the phantom. The standard approach using TOF- and deconvolution-based methods manages to identify the shape of the target, but the quality of the reconstruction is degraded by artefacts (Fig. \ref{fig:phantom_reconstruction}-B). Our proposed SRI approach manages to reduce the artefacts present in the reconstruction, increasing its quality and accuracy (Fig. \ref{fig:phantom_reconstruction}-C).

Table \ref{tab:phantom_reconstruction} contains quantitative metrics of the structural fidelity (SSIM) and reconstruction precision (SD) achieved in each of the three considered cases. These results show that the application of SRI not only reduces imaging artefacts, but also improves the quantitative accuracy and precision of the reconstruction.

\section{Discussion}

We have shown that SRI can produce experimental calibrations with higher accuracy than those generated with standard techniques, and that these improvements lead to reduced artefacts and increased reconstruction quality when applied to FWI imaging.

Compared to the baseline, the standard combination of TOF and deconvolution improves calibration accuracy, reducing error levels in phase and magnitude for all transducers. However, due to its design, the standard approach cannot adequately capture phase and amplitude variations at all emission and reception angles. In particular, the calibration is accurate directly in front of the transducer, i.e. when the angle between emitter and receiver is low, but becomes incorrect at larger angles, where the effect of the size and orientation of the transducers becomes more relevant.

Contrarily, SRI produces a robust calibration that remains highly accurate at all angles and for all transducers, both in transmission and in reception. Therefore, we conclude that SRI outperforms standard calibration methods, both in terms of magnitude and phase error. This is due to the fact that SRI is highly flexible, with its spatially distributed IR over the surface of the surrogate model being capable of complex manipulations of the wavefield that can absorb all linear sources of uncertainty present in our models of the acquisition device. However, the flexibility provided by SRI involves a higher computational cost than standard calibration methods. This is partly offset by the fact that calibrations remain usable for an extended period of time. Early results suggest that the performed calibrations remain stable and are unaffected by changes in surrounding conditions, such as the temperature of the water bath. Furthermore, the formulation of SRI means that the calibration can be carried out in arbitrarily complex media and that the calibration will be valid for different excitation signals as long as their frequency spectra overlap \cite{Cueto2021SpatialImaging}.

In order to produce successful SRI calibrations, it is important that the surrogate models are chosen such that the experimental transducers are fully enclosed by the hemispherical surface. Nonetheless, this could lead to increasingly large surfaces and a growing number of parameters if there are large deviations from the nominal locations. This could potentially make the calibration computationally prohibitive, but can be alleviated by ensuring that prior knowledge about the acquisition system is used to reduce location uncertainties before carrying out the calibration.

In this study, deconvolution was chosen as a comparison method to calibrate the IR of the transducers. This is because we showed in our previous work that deconvolution performed better than other commonly used calibration methods on average \cite{Cueto2021SpatialImaging}, while also being directly applicable both in transmission and in reception. Although an extension of TOF-like methods that includes information about the size of the transducer and its orientation could be formulated, the resulting optimisation problem would be computationally infeasible to solve. This is because it would require the explicit calculation of a sensitivity matrix or the use of global optimisation methods, which would have to deal with thousands of transducers to calibrate in practical acquisition devices. By contrast, SRI can effectively absorb size and orientation in a computationally efficient way thanks to its adjoint formulation. Our comparison in this study does not consider SRI in its original form because its practical application to calibrate the full acquisition system necessarily requires an approach that handles location and orientation as well as transducer IR.

Two points regarding our calibration setup require discussion. First, we were able to introduce the hydrophone vertically into the ring to perform our transmission calibration because of the small size of the needle (0.5 mm) with respect to the wavelength (approximately 3 mm at the centre frequency of the transducers), effectively rendering the hydrophone an almost omnidirectional point receiver. Second, the use of a smaller dataset in reception than in transmission had a limited impact on this particular calibration because of the simple geometry of the acquisition device. This smaller dataset was used because an equivalent omnidirectional point transmitter was not available during this study. We are actively exploring the use of such omnidirectional ultrasound transmitters in order to calibrate more complex acquisition geometries.

Our FWI reconstructions of the experimental phantom demonstrate that the use of SRI calibrations results in improved images, according to our quality control metrics, when compared to other calibration methods. The high accuracy achieved by the SRI calibration for all 40 transducers and across all angles suggests that the methodology is general in nature and could easily be applied to a wide range of acquisition geometries and imaging targets.

Some prior studies have been shown to be successful at imaging tissues such as the breast \cite{Wiskin20173-DResults,Sandhu2015FrequencyTransducer,CalderonAgudo20183DInversion} and the limbs \cite{Wiskin2020FullContrast} by using standard calibration methods. This is possible because these studies used transducers whose size was smaller than the dominant wavelength, or whose size could be considered small within a 2D imaging plane. In those cases, TOF-based methods can successfully estimate transducer locations, and deconvolution-like methods produce better IRs because the response of these smaller transducers is less angle-dependent than for larger ones. However, using smaller active surfaces results in lower signal-to-noise ratios, making it impossible to image high-contrast areas such as the adult human head. Additionally, 2D modelling and reconstruction are not applicable in the presence of complex, three-dimensional structures such as the skull \cite{Guasch2020Full-waveformBrain}. As we move towards the generalised application of ultrasound tomography techniques to complex tissues, the need for better calibration techniques that can reliably reproduce large transducers will increase and become an integral component of wave-equation based imaging algorithms

The biggest limitation of our proposed SRI methodology is the limited availability of calibration data in reception, which will degrade its performance when applied to more complex acquisition geometries. This can be solved by employing an omnidirectional source that can be used to produce a grid of measurements similar to that generated in transmission. This is the focus of ongoing research.

\section{Conclusion}

Recent advances in ultrasound computed tomography have showcased its potential to provide an affordable, portable and safe imaging modality that can produce high-resolution reconstructions of tissues that have long been considered out of reach for traditional ultrasound, such as the adult human brain. These techniques are orders of magnitude more computationally expensive than conventional ultrasound modalities, but are becoming closer to clinical feasibility thanks to progress in computational sciences coupled with increasingly lower hardware costs. However, these advances are hindered by the lack of methodologies to calibrate experimental tomographic acquisition devices beyond very specific configurations.

Consequently, we have introduced a technique that is capable of overcoming these limitations by extending our prior work on spatial response identification (SRI). Subsequently, we have experimentally demonstrated that the approach is flexible and robust, outperforming available standard procedures. This is possible because our approach reformulates the calibration problem, encoding the location, orientation and impulse response of the transducers into a black-box parametrisation while honouring the full wave equation. Our method presents the advantage that it can accurately recreate experimental transducer behaviour irrespective of emission and reception angles. This is a critical requirement for wave-based imaging methods due to their implicit assumption that wave propagation can be modelled with enough fidelity to match all experimental data. Additionally, we have proved that the application of SRI results in improved experimental full-waveform inversion reconstructions. 

SRI, then, offers a flexible solution that has the potential to enable the generalised experimental application of the new generation of ultrasound imaging modalities. Furthermore, other ultrasound applications that require accurate transducer modelling, such as high-intensity focused ultrasound or optoacoustic tomography, will also benefit from the flexibility and robustness that SRI provides.

\section*{Acknowledgement}

The authors are grateful to Thomas Robins for his assistance in manufacturing the tissue-mimicking phantom used in this study. 

\bibliographystyle{ieeetr}
\bibliography{main}

\end{document}